\documentclass[prd,aps,groupedaddress,showpacs,nofootinbib,amsmath]{revtex4}
\usepackage{epsf,epsfig,graphicx}

\newcounter{muni}

\begin{document}
\hbadness=10000 \pagenumbering{arabic}

\title{Pseudoscalar glueball mass from $\eta$-$\eta'$-$G$ mixing}

\author{Hai-Yang Cheng$^{1}$}
\email{phcheng@phys.sinica.edu.tw}
\author{Hsiang-nan Li$^{1,2,3}$}
\email{hnli@phys.sinica.edu.tw}
\author{Keh-Fei Liu$^{4}$}
\email{liu@pa.uky.edu  }

\affiliation{$^{1}$Institute of Physics, Academia Sinica, Taipei,
Taiwan 115, Republic of China}
\affiliation{$^{2}$Department of
Physics, Tsing-Hua University, Hsinchu, Taiwan 300, Republic of
China}
\affiliation{$^{3}$Department of Physics, National Cheng-Kung University,\\
Tainan, Taiwan 701, Republic of China}
\affiliation{$^{4}$Department of Physics and Astronomy, University
of Kentucky, Lexington, KY 40506}

\begin{abstract}

We deduce the mass of the pseudoscalar glueball $G$ from an
$\eta$-$\eta'$-$G$ mixing formalism based on the anomalous Ward
identity for transition matrix elements. With the inputs from the
recent KLOE experiment, we find a solution for the pseudoscalar
glueball mass around $(1.4\pm 0.1)$ GeV, which is fairly insensitive
to a range of inputs with or without Okubo-Zweig-Iizuka-rule
violating effects. This affirms that $\eta(1405)$, having a large
production rate in the radiative $J/\Psi$ decay and not seen in
$\gamma\gamma$ reactions, is indeed a leading candidate for the
pseudoscalar glueball. Other relevant quantities including the
anomaly and pseudoscalar density matrix elements are obtained. The
decay widths for $G\to \gamma\gamma$, $\ell^+\ell^-$ are also
predicted.

\bigskip

\end{abstract}


\maketitle

\section{INTRODUCTION}

The quest for pseudoscalar glueballs has continued for decades. An $E(1420)$
meson with a mass of 1426 MeV was first discovered at CERN in 1963
through $p\bar p$ interactions \cite{E(1420)}. In 1980, Mark II
observed that a $J/\psi$ meson decays via photon emission into a
resonance at a mass around 1440 MeV \cite{iota}. This new state,
named $\iota(1440)$ by Mark II and Crystal Ball Collaborations
\cite{CB}, was also once called $G(1440)$ in \cite{DJL81,MSC81}.
Shortly after the Mark II experiment, $E(1420)$ and $\iota(1440)$
were first proposed to be the pseudoscalar glueball candidates in
\cite{KI80} and in \cite{DJL81,MSC81,Lacaze}, respectively, while an
opposite opinion that $E(1420)$ was an $1^+$ $s\bar s$ quark state
was advocated in \cite{CEC81}. As the experimental situation was
sorted out, $E(1420)$ turned out to be an $1^+$ meson now known as
$f_1(1420)$, and $\iota(1440)$ was a pseudoscalar state now known as
$\eta(1405)$. For an excellent review of the $E$ and $\iota$ mesons,
see \cite{MCU06}.

$\eta(1405)$ indeed behaves like a glueball in its productions and
decays. The $K\bar K\pi$ and $\eta\pi\pi$ channels in $\gamma\gamma$
collisions have been investigated \cite{MA01}. While $\eta(1475)$ in
$K\bar K\pi$ was observed, $\eta(1405)$ in $\eta\pi\pi$ was not.
Since the glueball production is presumably suppressed in
$\gamma\gamma$ collisions, the above observations suggest that the
latter state has a large glueball content \cite{LYF03}. $J/\psi$
radiative decays through $\gamma gg$ have been considered as the
ideal channels of searching for glueballs. The branching ratio
${\cal B}(J/\psi \to \gamma \eta(1405))$ of order $10^{-3}$ is much
larger than the decays $J/\psi \to \gamma \eta(1295)$, $\gamma
\eta(2225)$,... which are either not seen or are of order $10^{-4}$.
The decay of a nearby $\eta(1475)\to\gamma\gamma$ has been observed
\cite{MA01}, but $\eta(1405)\to\gamma\gamma$ has not. All these
features support the proposal that $\eta(1405)$ is a good
pseudoscalar glueball candidate \cite{MCU06}. There were also
theoretical support based on the closed flux-tube model \cite{FNW04}
and the model that combines the octet, the singlet, and the glueball
into a decuplet \cite{GMM07}. Besides $\eta(1405)$, other states
with masses below 2 GeV have also been proposed as the candidates,
such as $\eta(1760)$ in \cite{WRZ00} and $X(1835)$ in \cite{KM05}.

As for the scalar glueball, two of the authors (HYC and KFL) and
Chua \cite{CCL} have considered a model for the glueball and $q\bar
q$ mixing, which involves the neutral scalar mesons $f_0(1370)$,
$f_0(1500)$ and $f_0(1710)$, based on two lattice results: (i) a
much better SU(3) symmetry in the scalar sector than in the other
meson sectors \cite{Mathur} and (ii) an unmixed scalar glueball at
about 1.7 GeV in the quenched approximation \cite{YC06}. It was
found that $f_0(1500)$ is a fairly pure octet, having very little
mixing with the singlet and the glueball, while $f_0(1370)$ and
$f_0(1710)$ are dominated by the glueball and the $q\bar q$ singlet,
respectively, with about $10\%$ mixing between them. The observed
enhancement of $\omega f_0(1710)$ production over $\phi f_0(1710)$
in hadronic $J/\psi$ decays and the copious $f_0(1710)$ production
in radiative $J/\psi$ decays lend further support to the prominent
glueball nature of $f_0(1710)$.

Contrary to the above case, the pseudoscalar glueball interpretation
for $\eta(1405)$ is, however, not favored by quenched lattice gauge
calculations, which predicted the mass of the $0^{-+}$ state to be
above 2 GeV in \cite{GSB93} and around 2.6 GeV in
\cite{Morningstar,YC06}. It is not favored by the sum-rule analysis
with predictions higher than 1.8 GeV \cite{sum,GG97} either. Readers
are referred to \cite{VKV08} for a recent review on the results of
the glueball masses. Note that the above lattice calculations were
performed under the quenched approximation without the fermion
determinants. It is believed that dynamical fermions may have a
significant effect in the pseudoscalar channel, because they raise
the singlet would-be-Goldstone boson mass from that of the pion to
$\eta$ and $\eta'$. It has been argued that the pseudoscalar
glueball mass in full QCD is substantially lower than that in the
quenched approximation \cite{GG97}. In view of the fact that the
topological susceptibility is large ($\approx (191 {\rm MeV})^4$) in
the quenched approximation \cite{Debbio}, and yet is zero for full
QCD in the chiral limit, it is conceivable that full QCD has a large
effect on the glueball as it does on $\eta$ and $\eta'$.

In this paper, we infer the pseudoscalar glueball mass $m_G$ from
the $\eta$-$\eta'$-$G$ mixing, where $G$ denotes the physical
pseudoscalar glueball. Implementing this mixing into the equations
of motion for the anomalous Ward identity, that connects the vacuum
to $\eta, \eta'$ and $G$ transition matrix elements of the
divergence of axial-vector currents to those of pseudoscalar
densities and the U(1) anomaly, $m_G$ is related to other
phenomenological quantities such as the $\eta, \eta'$ masses, the
decay constants, and the mixing angles. Since the mixing angles have
been measured recently from the $\phi \to \gamma\eta, \gamma\eta'$
decays by KLOE \cite{KLOE}, $m_G$ can be solved. Our numerical study
gives a fairly robust result $m_G \approx 1.4$ GeV, which is
insensitive to a range of inputs. We also obtain the matrix elements
for the pseudoscalar densities and axial U(1) anomaly associated
with the $\eta$, $\eta'$, and $G$ states. The values of the
pseudoscalar density matrix elements for the $\eta$, $\eta^{\prime}$
mesons are close to those obtained in the Feldmann-Kroll-Stech (FKS)
scheme \cite{FKS}, which does not include the mixing with the
pseudoscalar glueball. The results of the anomaly matrix elements
for the above states are quite consistent with those estimated from
the topological susceptibility \cite{Witten,Veneziano,Ohta2} and the
lattice evaluation \cite{YC06}, indicating that the $J/\psi\to\gamma
\eta'$ branching ratio could be comparable to that of
$J/\psi\to\gamma G$. We then study the pseudoscalar glueball decays
into two photons and two leptons $G\to \gamma\gamma$,
$\ell^+\ell^-$. The comparison of our analysis with the properties
of known mesons suggests that the $\eta(1405)$ meson is a strong
pseudoscalar glueball candidate.

In sec.~II we set up the formalism for the $\eta$-$\eta'$-$G$
mixing, assuming that the glueball only mixes with the
flavor-singlet $\eta_1$, but not with the flavor-octet $\eta_8$. Our
parametrization for the mixing matrix contains only two angles and
differs from that in \cite{KLOE}, where it is assumed that $\eta$
does not mix with the glueball state. The solution for the
pseudoscalar gluaball mass $m_G$ is derived in Sec.~III with the
phenomenological inputs from KLOE \cite{KLOE}. The solutions with
the inputs from \cite{EN07} and from \cite{FKS} as a limit of
vanishing mixture with the glueball state are also presented for
comparison. It will be shown in Sec.~IV that the result for $m_G$ is
stable against the variations of phenomenological inputs and of
corrections violating the Okubo-Zweig-Iizuka (OZI) rule \cite{OZI}.
The $G\to \gamma\gamma$, $\ell^+\ell^-$ decay widths are also
estimated. Section V is the conclusion.

\section{$\eta$-$\eta'$-$G$ MIXING}  \label{mixing}

We extend the FKS formalism \cite{FKS} for the $\eta$-$\eta'$ mixing
to include the pseudoscalar glueball $G$. In the FKS scheme, the
conventional singlet-octet basis and the quark-flavor basis have
been proposed. For the latter, the $q\bar q\equiv (u\bar u+d\bar
d)/\sqrt{2}$ and $s\bar s$ flavor states, labeled by the $\eta_q$
and $\eta_s$ mesons, respectively, are defined. In the extension to
the $\eta$-$\eta'$-$G$ mixing, the physical states $\eta$, $\eta'$
and $G$ are related to the octet, singlet, and unmixed glueball
states $\eta_8$, $\eta_1$ and $g$, respectively, through the
combination of rotations
\begin{equation}\label{qs}
   \left( \begin{array}{c}
    |\eta\rangle \\ |\eta'\rangle\\|G\rangle
   \end{array} \right)
   = U_3(\theta)U_1(\phi_G)
   \left( \begin{array}{c}
    |\eta_8\rangle \\ |\eta_1\rangle\\|g\rangle
   \end{array} \right) \;,
\end{equation}
with the matrices
\begin{equation}
U_3(\theta)=\left( \begin{array}{ccc}
    \cos\theta & -\sin\theta & 0\\
    \sin\theta & \cos\theta &0\\
    0 &0&1
   \end{array} \right)\;,\;\;\;\;
U_1(\phi_G)=\left( \begin{array}{ccc}
    1 &0 &0\\
    0 &\cos\phi_G & \sin\phi_G \\
    0 &-\sin\phi_G & \cos\phi_G
   \end{array} \right)\;.
\end{equation}
The matrix $U_1$ ($U_3$) represents a rotation around the axis along
the $\eta_8$ meson (unmixed glueball $g$). Equation~(\ref{qs}) is
based on the assumption that $\eta_8$ does not mix with the
glueball, under which two mixing angles $\theta$ and $\phi_G$ are
sufficient.

The octet and singlet states are related to the flavor states via
\begin{equation}
\left( \begin{array}{c}
    |\eta_8\rangle \\ |\eta_1\rangle\\|g\rangle
   \end{array} \right)
   = U_3(\theta_i)
\left( \begin{array}{c}
    |\eta_q\rangle \\ |\eta_s\rangle\\|g\rangle
   \end{array} \right) \;,
\end{equation}
where $\theta_i$ is the ideal mixing angle with
$\cos\theta_i=\sqrt{1/3}$ and $\sin\theta_i=\sqrt{2/3}$, i.e.,
$\theta_i=54.7^\circ$. The flavor states are then transformed into
the physical states through the mixing matrix
\begin{eqnarray}
U(\phi,\phi_G)&=&U_3(\theta)U_1(\phi_G)U_3(\theta_i)\;,\nonumber\\
&=&\left(
\begin{array}{ccc}
\cos\phi+\sin\theta\sin\theta_i\Delta_G & -\sin\phi+\sin\theta\cos\theta_i\Delta_G & -\sin\theta\sin\phi_G\\
\sin\phi-\cos\theta\sin\theta_i\Delta_G & \cos\phi-\cos\theta\cos\theta_i\Delta_G & \cos\theta\sin\phi_G\\
  -\sin\theta_i\sin\phi_G &-\cos\theta_i\sin\phi_G&\cos\phi_G
   \end{array} \right)\;,\label{mut}
\end{eqnarray}
with the angle $\phi=\theta+\theta_i$ and the abbreviation
$\Delta_G=1-\cos\phi_G$. $U$ has been written in the form, which
approaches the FKS mixing matrix \cite{FKS} in the $\phi_G\to 0$
limit. That is, the angle $\phi$ plays the same role as the mixing
angle in the FKS scheme.

Our formalism assumes isospin symmetry, i.e. no mixing with
$\pi^0$, and neglects other possible admixtures from $c\bar c$
states and radial excitations. The widely studied decay constants
$f_q$ and $f_s$ are defined by \cite{FKS}
\begin{eqnarray}
   \langle 0|\bar q\gamma^\mu\gamma_5 q|\eta_q(P)\rangle
   &=& -\frac{i}{\sqrt2}\,f_q\,P^\mu \;,\nonumber \\
   \langle 0|\bar s\gamma^\mu\gamma_5 s|\eta_s(P)\rangle
   &=& -i f_s\,P^\mu \;,\label{deffq}
\end{eqnarray}
for the light quark $q=u$ or $d$. The $\eta_q$ ($\eta_s$) meson
decay constant $f_q^s$ ($f_s^q$) through the $s$ ($q$) quark current
\cite{HCL08}, and the unmixed glueball decay constants $f_g^{q,s}$
through the $q$ and $s$ quark currents, can be defined in a similar
way:
\begin{eqnarray}
   & &\langle 0|\bar q\gamma^\mu\gamma_5 q|\eta_s(P),g(P)\rangle
   = -\frac{i}{\sqrt2}\,f_{s,g}^q\,P^\mu \;, \nonumber\\
   & &\langle 0|\bar s\gamma^\mu\gamma_5 s|\eta_q(P),g(P)\rangle
   = -i f_{q,g}^s\,P^\mu \;.
\label{offdiagonal}
\end{eqnarray}
The decay constants associated with the $\eta$ meson, $\eta'$ meson,
and the physical glueball defined in
\begin{eqnarray}\label{deffh}
   & &\langle 0|\bar q\gamma^\mu\gamma_5 q|\eta(P),\eta^{\prime}(P),G(P)\rangle
   = -\frac{i}{\sqrt2}\,f_{\eta,\eta^{\prime},G}^q\,P^\mu
   \;,\nonumber\\
   & &\langle 0|\bar s\gamma^\mu\gamma_5 s|\eta(P),\eta^{\prime}(P),G(P)\rangle
   = -i f_{\eta,\eta^{\prime},G}^s\,P^\mu \;,
\end{eqnarray}
are related to those associated with the $\eta_q$, $\eta_s$, and
$g$ states via the same mixing matrix
\begin{eqnarray}
\left(
\begin{array}{cc}
f_\eta^q & f_\eta^s \\
f_{\eta'}^q & f_{\eta'}^s \\
f_G^q &f_G^s
\end{array} \right)=
U(\phi,\phi_G) \left(
\begin{array}{cc}
f_q & f_q^s \\
f_s^q & f_s \\
f_g^q & f_g^s
\end{array} \right)
\;.\label{fpi}
\end{eqnarray}

Sandwiching the equations of motion for the anomalous Ward identity
\begin{eqnarray}
   \partial_\mu(\bar q\gamma^\mu\gamma_5 q) &=& 2im_q\,\bar q\gamma_5 q
   +\frac{\alpha_s}{4\pi}\,G_{\mu\nu}\,\widetilde{G}^{\mu\nu}\;,\nonumber\\
\partial_\mu(\bar s\gamma^\mu\gamma_5 s) &=& 2im_s\,\bar s\gamma_5s
   +\frac{\alpha_s}{4\pi}\,G_{\mu\nu}\,\widetilde{G}^{\mu\nu}\;,
   \label{eom}
\end{eqnarray}
between vacuum and $|\eta\rangle$, $|\eta'\rangle$ and $|G\rangle$, where
$G_{\mu\nu}$ is the field-strength tensor and
$\widetilde{G}^{\mu\nu}$ the dual field-strength tensor, and
following the procedure in \cite{HCL08}, we derive
\begin{eqnarray}
M_{qsg}^2=U^\dagger(\phi,\phi_G) M^2 U(\phi,\phi_G)\tilde
J\;.\label{matrix}
\end{eqnarray}
In the above expression the matrices are written as
\begin{eqnarray}
M_{qsg}^2&=&\left(\begin{array}{ccc}
m_{qq}^2+(\sqrt{2}/f_q)\langle 0|\alpha_sG{\tilde
G}/(4\pi)|\eta_q\rangle & m_{sq}^2+(1/f_s)\langle
0|\alpha_sG{\tilde G}/(4\pi)|\eta_q\rangle & 0\\
              m_{qs}^2+(\sqrt{2}/f_q)\langle
0|\alpha_sG{\tilde G}/(4\pi)|\eta_s\rangle &
m_{ss}^2+(1/f_s)\langle 0|\alpha_sG{\tilde G}/(4\pi)|\eta_s\rangle
&0
\\
m_{qg}^2+(\sqrt{2}/f_q)\langle 0|\alpha_sG{\tilde
G}/(4\pi)|g\rangle & m_{sg}^2+(1/f_s)\langle 0|\alpha_sG{\tilde
G}/(4\pi)|g\rangle &0
\end{array}\right)\;,\nonumber\\
M^2&=&\left(\begin{array}{ccc}
  m_{\eta}^2 & 0 &0\\
  0 & m_{\eta'}^2&0 \\
  0 &0 & m_G^2
\end{array} \right)\;,\;\;\;\;
\tilde J=\left(\begin{array}{ccc}
   1 & f_q^s/f_s &0\\
  f_s^q/f_q & 1 & 0 \\
  f_g^q/f_q &f_g^s/f_s & 0
\end{array} \right)\;,\label{I}
\end{eqnarray}
with the abbreviation
\begin{eqnarray}
m_{qq,qs,qg}^2&\equiv&\frac{\sqrt{2}}{f_q}\langle 0|m_u\bar u i\gamma_5
u+m_d\bar d
i\gamma_5 d|\eta_q,\eta_s,g\rangle\;,\nonumber\\
m_{sq,ss,sg}^2&\equiv&\frac{2}{f_s}\langle 0|m_s\bar s i\gamma_5
s|\eta_q,\eta_s,g\rangle\;.\label{mqq}
\end{eqnarray}

In the limit of the large color number $N_c$, the scaling for the
decay constants, the pseudoscalar densities, and the anomaly matrix
elements is \cite{tHooft}
\begin{eqnarray} \label{largeNc}
&& \qquad\qquad  f_{q,s} \sim  O(\sqrt{N_c})\;, \qquad
f_g^{q,s}\sim O(1)\;, \qquad f_q^s\sim f_s^q\sim
O(1/\sqrt{N_c})\;,
\nonumber \\
&& \qquad\qquad  m_G \sim O(1), \qquad\quad\quad
\phi_G\sim O(1/\sqrt{N_c})\;,\nonumber\\
&& \qquad\qquad  m_{qq}^2 \sim O(1), \qquad\quad\quad m_{ss}^2\sim
O(1), \qquad \nonumber\\
&&\qquad\qquad m_{qg}^2\sim m_{sg}^2 \sim
O(1/\sqrt{N_c})\;,\qquad\quad\quad\qquad
m_{qs}^2\sim m_{sq}^2 \sim O(1/N_c)\;,   \nonumber\\
&& \langle 0|\alpha_sG{\tilde G}/(4\pi)|g\rangle \sim O(1), \qquad
\langle 0|\alpha_sG{\tilde G}/(4\pi)|\eta_q\rangle \sim \langle
0|\alpha_sG{\tilde G}/(4\pi)|\eta_s\rangle \sim O(1/\sqrt{N_c})\;.
\end{eqnarray}
The pseudoscalar meson and glueball masses scale as $O(1)$ in
large $N_c$. However, it has been pointed out
\cite{Witten,Veneziano,Ohta2} that the sub-leading $O(1/N_c)$ term
in the $\eta_{1}$ mass squared $m_{\eta_{_1}}^2\sim O(1)+O(1/N_c)$
is numerically large due to the $U(1)$ anomaly, and is related to
the topological susceptibility $\chi$ in the quenched QCD without
fermions. In the chiral limit, the relation $m_{\eta'}^2 = 4 N_F
\chi/f_{\pi}^2 = 2 \sqrt{N_F}\langle 0|\alpha_sG{\tilde
G}/(4\pi)|\eta'\rangle/f_{\pi}$ with $N_F$ being the number of
flavors gives $\langle 0|\alpha_sG{\tilde G}/(4\pi)|\eta'\rangle
\approx 0.035\,{\rm GeV^3}$ for $\chi = (191\, \rm{MeV})^4$
\cite{Debbio}. Although being $O(1/\sqrt{N_c})$ in large $N_c$,
this anomaly matrix element is numerically much larger than the
$O(\sqrt{N_c})$ quantities $m_{qq}^2 f_q \approx 0.0026\,
\rm{GeV}^3$ for $m_{qq}^2 \approx m_{\pi}^2$ and comparable to
$m_{ss}^2 f_s \approx 0.087\, \rm{GeV}^3$ for $m_{ss}^2 \approx 2
m_K^2 - m_{\pi}^2$. In view of this, we shall keep all the anomaly
matrix elements for $\eta_q, \eta_s$ and $g$ in the following
analysis. On the other hand, we expect the decay constants and the
pseudoscalar density matrix elements to have the normal ordering
in terms of $N_c$. That is, we expect $f_{q,s} > f_g^{q,s} >
f_q^s, f_s^q$, $m_{qq}^2
> m_{qg}^2 > m_{qs}^2$, and $m_{ss}^2 > m_{sg}^2  > m_{sq}^2$.
The above ordering is consistent with the OZI rule in that double
quark annihilation, which is present in $f_q^s$, $f_s^q$, $m_{qs}^2$
and $m_{sq}^2$ but not in others, is OZI-rule violating and
suppressed. We note that the two sides of each of the equations in
Eq.~(\ref{matrix}) have the same $N_c$ scaling, implying consistency
of our formalism in terms of $N_c$.

\section{PSEUDOSCALAR GLUEBALL MASS}  \label{glueball}

The explicit expansion of Eq.~(\ref{matrix}) leads to
\begin{eqnarray}
&& U_{11}^\dagger[U_{11}+U_{12}R'+U_{13}r']m_{\eta}^2+
U_{12}^\dagger[U_{21}+U_{22}R'+U_{23}r']m_{\eta'}^2+
U_{13}^\dagger[U_{31}+U_{32}R'+U_{33}r']m_G^2 \nonumber\\
&& \qquad =m_{qq}^2+(\sqrt{2}/f_q)\langle
0|\alpha_sG{\tilde G}/(4\pi)|\eta_q\rangle\;, \label{1m}\\
&& U_{11}^\dagger[U_{11}R+U_{12}+U_{13}r]m_{\eta}^2+
U_{12}^\dagger[U_{21}R+U_{22}+U_{23}r]m_{\eta'}^2+
U_{13}^\dagger[U_{31}R+U_{32}+U_{33}r]m_G^2 \nonumber\\
&& \qquad =m_{sq}^2+(1/f_s)\langle
0|\alpha_sG{\tilde G}/(4\pi)|\eta_q\rangle\;,\label{2m}\\
& &  \qquad =m_{qs}^2+(\sqrt{2}/f_q)\langle 0|\alpha_sG{\tilde
G}/(4\pi)|\eta_s\rangle\;,\label{3m}\\
&& U_{21}^\dagger[U_{11}R+U_{12}+U_{13}r]m_{\eta}^2+
U_{22}^\dagger[U_{21}R+U_{22}+U_{23}r]m_{\eta'}^2+
U_{23}^\dagger[U_{31}R+U_{32}+U_{33}r]m_G^2 \nonumber\\
& & \qquad =m_{ss}^2 +(1/f_s)\langle 0|\alpha_sG{\tilde
G}/(4\pi)|\eta_s\rangle\;,\label{4m}\\
&& U_{31}^\dagger[U_{11}+U_{12}R'+U_{13}r']m_{\eta}^2+
U_{32}^\dagger[U_{21}+U_{22}R'+U_{23}r']m_{\eta'}^2+
U_{33}^\dagger[U_{31}+U_{32}R'+U_{33}r']m_G^2 \nonumber\\
& &  \qquad =m_{qg}^2+(\sqrt{2}/f_q)\langle 0|\alpha_sG{\tilde
G}/(4\pi)|g\rangle\;,\label{5m}\\
&& U_{31}^\dagger[U_{11}R+U_{12}+U_{13}r]m_{\eta}^2+
U_{32}^\dagger[U_{21}R+U_{22}+U_{23}r]m_{\eta'}^2+
U_{33}^\dagger[U_{31}R+U_{32}+U_{33}r]m_G^2 \nonumber\\
& & \qquad=m_{sg}^2+(1/f_s)\langle 0|\alpha_sG{\tilde
G}/(4\pi)|g\rangle\;,\label{6m}
\end{eqnarray}
where the parameters $r\equiv f_g^s/f_s$, $r'\equiv f_g^q/f_q$, $R
\equiv f_q^s/f_s$, and $R'\equiv f_s^q/f_q$ are introduced, and
$U_{ij}$ denotes the matrix element of $U$. In developing our mixing
formalism, the flavor-independent couplings between the glueball $g$
and the pseudoscalar $u\bar{u}, d\bar{d}$ and $s\bar{s}$ states are
assumed, so that $g$ only mixes with the flavor-singlet $\eta_1$.
Accordingly, we postulate $f_g^q = \sqrt{2} f_g^s$ and $f_s^q =
f_q^s$, and thus the relations
\begin{eqnarray}
r'=\sqrt{2}{f_s\over f_q}r \qquad R'={f_s\over f_q}R, \label{ratio}
\end{eqnarray}
which will be adopted in the numerical study in Sec.~IV.

We first explore the implication of the $\eta$-$\eta'$-$G$ mixing on
the glueball mass $m_G$. To simplify the matter, the ratios $r$,
$r'$, $R$ and $R'$ are neglected, which are $O(1/\sqrt{N_c})$ and
$O(1/N_c)$, respectively, in large $N_c$ as shown in
Eq.~(\ref{largeNc}). We also neglect $m_{qg}^2$ and $m_{sg}^2$
relative to the numerically large anomaly term $\langle
0|\alpha_sG{\tilde G}/(4\pi)|g\rangle/f_{q,s}$ as an approximation.
It should be safe to drop $m_{qg}^2$, since it is, like the small
$m_{qq}^2\approx m_{\pi}^2$, proportional to the light $u/d$ quark
mass. On the other hand, it is not clear if it is safe to drop
$m_{sg}^2$. Although it is $O(1/\sqrt{N_c})$ compared to $m_{ss}^2$,
but the latter, being proportional to the strange quark mass, is
larger than $\langle 0|\alpha_sG{\tilde G}/(4\pi)|\eta'\rangle/f_s$
in the chiral limit as discussed in Sec.~\ref{mixing}. This subject
will be investigated in a more detailed numerical analysis later in
Sec.~\ref{NA}. Having made the above assumptions, we take the ratio
of Eqs.~(\ref{5m}) and (\ref{6m}), and obtain
\begin{eqnarray}
\frac{c\theta (s\phi-c\theta s\theta_i\Delta_G)m_{\eta'}^2-s\theta
(c\phi+s\theta s\theta_i\Delta_G)^2m_\eta^2 -s\theta_i c\phi_G
m_G^2}{c\theta (c\phi-c\theta c\theta_i\Delta_G)m_{\eta'}^2+s\theta
(s\phi-s\theta c\theta_i\Delta_G)^2m_\eta^2 -c\theta_i c\phi_G
m_G^2}=\frac{\sqrt{2}f_s}{f_q}\;,\label{rg}
\end{eqnarray}
where $c\phi$ ($s\phi$) is the shorthand notation for $\cos\phi$
($\sin\phi$) and similarly for others.

Note that the above simple formula still holds, even after keeping
the $r'$- and $r$-dependent terms, as long as they obey
Eq.~(\ref{ratio}). In other words, the $r'$-dependent terms in Eq.~(\ref{5m}) and
the $r$-dependent terms in Eq.~(\ref{6m}) can be absorbed into the
right-hand sides of these equations and are therefore canceled after
taking the ratio of Eqs.~(\ref{5m}) and (\ref{6m}). The factor
$\sin\phi_G$ in the numerator and the denominator of Eq.~(\ref{rg})
has been canceled out, so that the $\phi_G$ dependence appears at
order of $\Delta_G\approx \phi_G^2$ for small $\phi_G$. As such, we
find that the solution for $m_G$ is stable against the most
uncertain input $\phi_G$, as long as the $\eta, \eta'$ mesons do not
mix with the glueball $g$ intensively. The solution depends on the
ratio $f_s/f_q$, which is, through Eqs.~(\ref{2m}) and (\ref{3m}),
related to
\begin{eqnarray}
\frac{\sqrt{2}f_s}{f_q}=\frac{\langle 0|\alpha_sG{\tilde{G}}/(4\pi)|\eta_q\rangle}
{\langle 0|\alpha_sG{\tilde{G}}/(4\pi)|\eta_s\rangle}\;,
\end{eqnarray}
if the pseudoscalar density matrix elements $m_{qs}^2$ and
$m_{sq}^2$ are neglected. It implies that the SU(3) symmetry
breaking in the axial anomaly matrix element echoes the symmetry
breaking in the decay constants, and plays a sensitive role in the
determination of the pseudoscalar glueball mass. For a given
$\phi_G$, $m_G$ increases with decreasing $f_s/f_q$. By the same
token, when the anomaly matrix element for the $\eta_s$ becomes
larger relative to that of the $\eta_q$ meson, the mass of the
pseudoscalar glueball gets higher.

Before solving for $m_G$ from Eq.~(\ref{rg}), we explain the
strategies for data fitting adopted in \cite{KLOE} and \cite{EN07},
which led to different extractions of the mixing angle $\phi_G$. In
Ref.~\cite{KLOE}, the decay constants $f_q=(1\pm 0.01)f_\pi$ and
$f_s=(1.4\pm 0.014)f_\pi$ \cite{TF00}, and the parameters associated
with meson wave function overlaps \cite{BES01} were fixed as inputs.
The angles $\phi=(39.7\pm 0.7)^\circ$ and $\phi_G=(22\pm 3)^\circ$
were then determined from the relevant data. A tiny error (1\%) was
assigned to $f_q$ and $f_s$, which is one of the reasons why a high
precision was reached for the determination of $\phi_G$. In
\cite{EN07} the data of $P\to \gamma V$ and $V\to \gamma P$ were
first considered, which do not depend on $f_q$ and $f_s$, and the
fit gave the outcomes $\phi=(41.4\pm 1.3)^\circ$ and $\phi_G=(12\pm
13)^\circ$. Without precise inputs of $f_q$ and $f_s$, and with the
hadronic parameters for meson wave function overlaps being free, it
is not unexpected to get a wide range for $\phi_G$. The value of
$\phi_G$, being consistent with zero~\cite{CH81}, means that the
data could be accommodated by the hadronic uncertainty alone. The
extracted $\phi$ and $\phi_G$ were then used as inputs to determine
$f_q$ and $f_s$ from the $\eta, \eta' \to \gamma\gamma$ data. Since
$\phi_G$ has a wide range, the results $f_q=(1.05\pm 0.03)f_\pi$ and
$f_s= (1.57\pm 0.28)f_\pi$ also have larger errors. The correlation
between $\phi_G$ and $f_s$ (a smaller $\phi_G$ corresponding to a
larger $f_s$) is a consequence of the constraint from these data. We
also note that a larger mixing angle $\phi_G=(33\pm 13)^\circ$ has
been extracted from the $J/\psi\to VP$ data recently \cite{E08}. In
summary, both sets of parameters in \cite{KLOE,EN07} can fit the
data, and are consistent with each other within their uncertainties.
It is seen that $f_q$, $f_s$, and $\phi$ are more or less certain,
but $\phi_G$ varies in a wider range. Fortunately, the solution for
the pseudoscalar glueball mass $m_G$ is not sensitive to $\phi_G$ as
discussed above and will be explored further in the remainder of
this paper.

As stated before, KLOE postulated that the glueball does not mix
with $\eta$ \cite{KLOE}. We shall point out that this postulate does
not yield a solution for $m_G$ in our formalism. The KLOE
parametrization for the $\eta$-$\eta'$-$G$ mixing matrix is written
as
\begin{eqnarray}
U_{\rm KLOE}&=&\left(
\begin{array}{ccc}
\cos\phi & -\sin\phi  & 0\\
\sin\phi\cos\phi_G & \cos\phi\cos\phi_G & \sin\phi_G\\
-\sin\phi\sin\phi_G &-\cos\phi\sin\phi_G&\cos\phi_G
   \end{array} \right)\;.\label{mKLOE}
\end{eqnarray}
Repeating the above procedure, Eq.~(\ref{rg}) is modified to
\begin{eqnarray}
\frac{s\phi c\phi_Gs\phi_G(m_{\eta'}^2-m_G^2)}{c\phi
c\phi_Gs\phi_G(m_{\eta'}^2-m_G^2)}=\frac{s\phi}{c\phi}=\frac{\sqrt{2}f_s}{f_q}\;.\label{rg2}
\end{eqnarray}
For the KLOE parameter set $f_q=f_\pi$, $f_s=1.4f_\pi$ and
$\phi=39.7^\circ$ \cite{KLOE}, Eq.~(\ref{rg2}) does not hold, and
there is no solution for $m_G$ as a result.

Since we have changed the mixing matrix from KLOE's in
Eq.~(\ref{mKLOE}) to Eq.~(\ref{mut}), we need to refit $\phi$ and
$\phi_G$ in principle. However, comparing Eqs.~(\ref{mKLOE}) and
(\ref{mut}), it is easy to find that their $2\times 2$ sub-matrices
in the left-upper hand corner have almost equal elements for
$\phi\approx 40^\circ$ and $\phi_G\approx 22^\circ$:
\begin{eqnarray}
U=\left(\begin{array}{ccc}
  0.751 & -0.654 &0.097\\
  0.585 & 0.725 & 0.362 \\
  -0.306 &-0.216 & 0.927
\end{array} \right)\;,\;\;\;\;
U_{\rm KLOE}=\left(\begin{array}{ccc}
  0.766 & -0.643 &0\\
  0.596 & 0.710 & 0.375 \\
  -0.241 &-0.287 & 0.927
\end{array} \right)\;.\label{com}
\end{eqnarray}
These four elements, which are responsible for the quark mixing, are
the only ones involved in the data fitting of $\phi \to\gamma\eta,
\gamma\eta'$, $\eta' \to \gamma\rho,\gamma\omega$, and $\eta,\eta'
\to \gamma\gamma$ mentioned above. Therefore, it is expected that
the refit of the data using our parametrization will give the mixing
angles close to KLOE's. That is, the KLOE parameter set can be
employed directly in our numerical analysis within uncertainty. The
other off-diagonal elements in Eq.~(\ref{com}), describing the
mixing among the $\eta, \eta'$ mesons and the glueball, do have
different values. It is thus understood why the two
parametrizations have similar mixing angles, but the ratios in
Eqs.~(\ref{rg}) and (\ref{rg2}) exhibit different behaviors as far
as $m_G$ is concerned.

It is also interesting to consider the parameter set from \cite{FKS}
with $f_q=(1.07\pm 0.02)f_\pi$, $f_s=(1.34\pm 0.06)f_\pi$,
$\phi=(39.3\pm 1.0)^\circ$, and $\phi_G = 0$ (no mixing with the
pseudoscalar glueball). Note that the lower $f_s$ in \cite{FKS}
arises from combined experimental and phenomenological constraints.
If  only the experimental constraints were used, mainly those of the
$\eta, \eta' \to\gamma\gamma$ data, its central value would increase
and the range is enlarged, giving $f_s=(1.42\pm 0.16) f_\pi$ close
to that extracted in \cite{KLOE}. Using the central values of
$f_s/f_q$ and $\phi_G$ from \cite{KLOE,FKS,EN07} as inputs, we
derive the pseudoscalar glueball mass from Eq.~(\ref{rg}) [see also
Eqs. (\ref{sol.3})-(\ref{solfks}) below]
\begin{eqnarray}
m_G=1.41, \quad 1.56, \quad 1.30\;\;{\rm GeV}\;, \label{solmg}
\end{eqnarray}
respectively. The above investigation leads to $m_G=(1.4\pm 0.1)$
GeV with the currently determined phenomenological
parameters. The proximity of the predicted $m_G$ to the mass of
$\eta(1405)$ and other properties of $\eta(1405)$ make it a very
strong candidate for the pseudoscalar glueball. We shall come back
to visit the robustness of our prediction in the next section, when higher
order effects in $1/N_c$ are included .

One may question whether other pseudoscalar mesons with masses
around 1.4 GeV, such as $\eta(1295)$ and $\eta(1475)$, should be
included into our mixing formalism. We note that $\eta(1295)$ and
$\eta(1475)$ have been assigned as the $2^1S_0$ states, namely, the
radial excitations of $q\bar q$ and $s\bar s$, respectively
\cite{PDG,GLT08}. As stated in the previous section, these radial
excitations do not mix with the $\eta$ and $\eta'$ mesons by
definition, since they are diagonalized under the same Hamiltonian.
As for the mixing of the radial excitations with the glueball, we
speculate that it is negligible for the following reason.
$\eta(1295)$ is practically degenerate with the radially excited
pion $\pi(1300)$, and $\eta(1475)$ is about 200 MeV above
$\eta(1295)$, a situation similar to the ideal mixing in the vector
meson sector with $\phi(1020)$ being $\sim 200$ MeV above
$\omega(780)$. This suggests that $\eta(1295)$ and $\eta(1475)$ are
much like the radially excited isovector pseudoscalar $q\bar q$ and
pseudoscalar $s\bar{s}$ without annihilation. The difference between
the isoscalar $\eta$, $\eta'$ mesons and the isovector pion is that
the former have disconnected insertions (annihilation) with the
coupling going through the contact term in the topological
susceptibility which pushes their masses up. By virtue of the fact
that $\eta(1295)$ is degenerate with $\pi(1300)$ and $\eta(1475)$ is
$\sim 200$ MeV above, they  do not seem to acquire such an
enhancement for their masses. Therefore, we venture to suggest that
the annihilation process is not important for these two mesons, and
their mixing with the glueball is weak.

A pseudoscalar glueball mass about $1.4$ GeV was also determined
from the framework of the $\eta$-$\eta'$-$G$ mixing in
\cite{LZC07}, but with a strategy quite different from ours: The
mixing is assumed to occur through a perturbative potential, so that
the mixing angles are parametrized in terms of the transition
strength among the states $\eta_q$, $\eta_s$ and $g$ and their
masses $m_{\eta_q}$, $m_{\eta_s}$ and $m_g$ \cite{LZC07}.
These parameters were then
fixed from data fitting. Hence, it is the unmixed glueball mass
$m_g$, instead of the physical glueball mass $m_G$, that was
derived in \cite{LZC07}. Moreover, the result of \cite{LZC07} is a
consequence of data fitting, while ours comes from the
solution to Eq.~(\ref{rg}). If the quark flavor states do not mix
strongly with the glueball, $m_G$ is expected to be close to that
of $m_g$. Following this reasoning, three possible $0^{-+}$
glueball candidates, $\eta(1405)$, $\eta(1475)$, and $X(1835)$
with masses around 1.4 GeV, have been examined in \cite{LZC07},
and the latter two were found to be experimentally disfavored.

\section{NUMERICAL ANALYSIS}  \label{NA}

We now proceed to solve Eqs.~(\ref{1m})-(\ref{6m}) based on the
large $N_c$ hierarchy in Eq.~(\ref{largeNc}). As discussed in
Sec.~\ref{mixing}, all the anomaly matrix elements will be kept.
Even though they are small parametrically ($O(1)$ and
$O(1/\sqrt{N_c})$), they are large numerically. As a follow up of
the last section, we first neglect the decay constants $f_q^s$,
$f_s^q$, and $f_g^{q,s}$, which are $O(1/\sqrt{N_c})$ and $O(1/N_c)$
lower than $f_{q,s}$, respectively. We also neglect the pseudoscalar
density matrix elements $m_{qg}^2$, $m_{sg}^2$, $m_{qs}^2$, and
$m_{sq}^2$, which are similarly suppressed as compared to $m_{qq}^2$
and $m_{ss}^2$\footnote{The off-diagonal mass terms $m_{sq}^2$ and
$m_{qs}^2$ have been absorbed into the matrix elements $\langle
0|\alpha_sG{\tilde G}/(4\pi)|\eta_q\rangle$ and $\langle
0|\alpha_sG{\tilde G}/(4\pi)|\eta_s\rangle$, respectively in
\cite{FKS}.}. Under this approximation, our formalism involves six
unknowns: three mass related terms $m_G$, $m_{qq}^2$, and
$m_{ss}^2$, and three anomaly matrix elements $\langle
0|\alpha_sG{\tilde G}/(4\pi)|\eta_q\rangle$, $\langle
0|\alpha_sG{\tilde G}/(4\pi)|\eta_s\rangle$, and $\langle
0|\alpha_sG{\tilde G}/(4\pi)|g\rangle$, provided that the
phenomenological quantities $m_{\eta}^2$, $m_{\eta'}^2$, $f_q$,
$f_s$, $\phi$ and $\phi_G$ are given as inputs. There are six
equations from Eq.~(\ref{matrix}), so the six unknowns can be solved
in principle. We note in passing that the four unknowns $m_{qq}^2$,
$m_{ss}^2$, $\langle 0|\alpha_sG{\tilde G}/(4\pi)|\eta_q\rangle$,
and $\langle 0|\alpha_sG{\tilde G}/(4\pi)|\eta_s\rangle$ were solved
for five given inputs $m_{\eta}^2$, $m_{\eta'}^2$, $f_q$, $f_s$, and
$\phi$ in the $\eta$-$\eta'$ mixing case \cite{CKL06} .

Using the central values of the parameter sets from \cite{KLOE},
\cite{EN07}, and \cite{FKS} for $f_q, f_s, \phi$ and $\phi_G$ as inputs,
we obtain the following solutions
\begin{eqnarray}
&m_{qq}^2=-0.073\;\;{\rm GeV}^2\;, &\langle 0|\alpha_sG{\tilde
G}/(4\pi)|\eta_q\rangle=0.069\;\;{\rm GeV}^3\;, \nonumber\\
&m_{ss}^2=0.52\;\;{\rm GeV}^2\;, &\langle 0|\alpha_sG{\tilde
G}/(4\pi)|\eta_s\rangle=0.035\;\;{\rm GeV}^3\;,\nonumber\\
&m_G=1.41\;\;{\rm GeV}\;, &\langle 0|\alpha_sG{\tilde
G}/(4\pi)|g\rangle=-0.033\;\;{\rm GeV}^3\;, \label{sol.3}
\end{eqnarray}
\begin{eqnarray}
&m_{qq}^2=-0.084\;\;{\rm GeV}^2\;, &\langle 0|\alpha_sG{\tilde
G}/(4\pi)|\eta_q\rangle=0.067\;\;{\rm GeV}^3\;,\nonumber\\
&m_{ss}^2=0.50\;\;{\rm GeV}^2\;, &\langle 0|\alpha_sG{\tilde
G}/(4\pi)|\eta_s\rangle=0.032\;\;{\rm GeV}^3\;,\nonumber\\
&m_G=1.30\;\;{\rm GeV}\;, &\langle 0|\alpha_sG{\tilde
G}/(4\pi)|g\rangle=-0.015\;\;{\rm GeV}^3\;, \label{solen}
\end{eqnarray}
and
\begin{eqnarray}
&m_{qq}^2=0.012\;\;{\rm GeV}^2\;, &\langle 0|\alpha_sG{\tilde
G}/(4\pi)|\eta_q\rangle=0.054\;\;{\rm GeV}^3\;,\nonumber\\
&m_{ss}^2=0.50\;\;{\rm GeV}^2\;, &\langle 0|\alpha_sG{\tilde
G}/(4\pi)|\eta_s\rangle=0.030\;\;{\rm GeV}^3\;,\nonumber\\
&m_G=1.56\;\;{\rm GeV}\;, &\langle 0|\alpha_sG{\tilde
G}/(4\pi)|g\rangle=0\;\;{\rm GeV}^3\;. \label{solfks}
\end{eqnarray}

The above solutions give us an idea of the range of uncertainties in
our predictions. It is observed that the solutions for the anomaly
matrix elements associated with the $\eta_q$ and $\eta_s$ mesons
change little in Eqs.~(\ref{sol.3})-(\ref{solfks}). However,
$\langle 0|\alpha_sG{\tilde G}/(4\pi)|g\rangle$ for the pseudoscalar
glueball varies with the inputs of $\phi_G$ and $f_s/f_q$ strongly
as can be seen from Eqs.~(\ref{5m}) and (\ref{6m}). The solutions of
$m_{qq}^2$ in Eqs.~(\ref{sol.3}) and (\ref{solen}) deviate from the
naive expectation $m_{qq}^2=m_\pi^2$ \cite{FKS}, while that in
Eq.~(\ref{solfks}) is in better agreement with $m_\pi^2$ due to a
smaller $f_s$. The solutions of $m_{ss}^2$, on the other hand, are
stable with respect to the various inputs, and are close to the
expected leading $N_c$ result $m_{ss}^2=2m_K^2-m_\pi^2$. The values
for $m_G$ have been shown in Eq.~(\ref{solmg}) already. We should
comment that $m_{qq}^2$ is small because of the cancellation of two
large terms as pointed out in \cite{CKL06}. It flips sign easily,
depending on the inputs of $f_s/f_q$ and OZI-rule violating effects,
which have been considered before in the two-angle formalism for the
$\eta$-$\eta'$ mixing \cite{SSW,HL98}. Our opinion is that
introducing the OZI-rule suppressed decay constants $f_q^s$, $f_s^q$
\cite{HCL08} is more transparent than employing the multiple-angle
formalism. It has been observed that the tiny corrections from
$f_q^s$ and $f_s^q$ can turn a negative $m_{qq}^2$ into a positive
value easily due to the smallness of $m_{qq}^2$ \cite{HCL08}.

In the following, we investigate the higher $O(1/N_c)$ effects from
the decay constants $f_g^{q,s}$, $f_q^s$, and $f_s^q$, i.e., from
$r$ and $R$ on our solutions. The pseudoscalar density matrix
elements $m_{qg}^2$, $m_{sg}^2$, $m_{qs}^2$, and $m_{sq}^2$ are
still ignored. We take $m_{qq}^2=m_\pi^2$, $m_{ss}^2=2m_K^2-m_\pi^2$
and $\langle 0|\alpha_sG{\tilde G}/(4\pi)|\eta_q\rangle=0.065$,
0.050, and 0.035 ${\rm GeV}^3$ (the typical values from
Eqs.~(\ref{sol.3})-(\ref{solfks})), as the inputs in order to solve
for the unknowns $r$, $R$ and $f_s$. The relation $m_{ss}^2=
2m_K^2-m_\pi^2$ seems to hold well for the earlier solutions in
Eqs.~(\ref{sol.3})-(\ref{solfks}). Thus, it is reasonable to fix it
to its leading $N_c$ value. Taking $f_q=f_\pi$, $\phi=42.4^\circ$
and $\phi_G=22^\circ$ and $12^\circ$, the corresponding solutions
are listed in Table~\ref{tab:1}. The results of $R$, $f_s$, and
$m_G$ are independent of the inputs of $\langle 0|\alpha_sG{\tilde
G}/(4\pi)|\eta_q\rangle$, reaffirming that $m_G$ is independent of
$r$ as seen from Eq.~(\ref{rg}). The magnitude of $R$ is smaller
than that of $r$, which in turn is smaller than unity. This finding
is in agreement with the large $N_c$ counting rule. The decay
constant $f_s$ turns out to be lower than those in \cite{KLOE,EN07},
following from the observation that a smaller $f_s$ leads to a
positive $m_{qq}^2$ \cite{CKL06}. The values of $m_G$ and $\langle
0|\alpha_sG{\tilde G}/(4\pi)|\eta_s\rangle$ are consistent with the
range derived in Eqs.~(\ref{sol.3})-(\ref{solfks}), implying that
these higher $O(1/N_c)$ effects are small. The parameters $r$ and
$\langle 0|\alpha_sG{\tilde G}/(4\pi)|g\rangle$ are found to be
sensitive to the inputs, and both of them increase with decreasing
$\langle 0|\alpha_sG{\tilde G}/(4\pi)|\eta_q\rangle$.

\begin{table}[t]
\caption{Solutions for various inputs of $\langle
0|\alpha_sG{\tilde G}/(4\pi)|\eta_q\rangle=0.065$ GeV$^3$ (the first
row), 0.050 GeV$^3$ (the second row), and 0.035 GeV$^3$ (the third
row) with $m_{qq}^2=m_\pi^2$, $m_{ss}^2=2m_K^2-m_\pi^2$,
$m_{sg}^2=m_{qg}^2=m_{qs}^2=m_{sq}^2=0$ and $\phi=42.4^\circ$. The upper
(lower) table is for $\phi_G=22^\circ$ ($\phi_G=12^\circ$).}
\begin{tabular}{ccc c c c} \toprule
$r$ & $R$ & $f_s$ & $m_G$\,(GeV) & $\langle
0|{\alpha_s\over 4\pi}G\tilde{G}|\eta_s\rangle$\,(GeV$^3$)
& $\langle 0|{\alpha_s\over 4\pi}G\tilde{G}|g\rangle$\,(GeV$^3$) \\
\colrule
$0.004$ & $-0.002$ & $1.25f_\pi$  & $1.50$ & 0.037 & $-0.038$   \\
$0.22$ & $-0.002$ & $1.25f_\pi$  & $1.50$ & 0.028 & 0.036   \\
$0.44$ & $-0.002$ & $1.25f_\pi$  & $1.50$ & 0.020 & 0.111   \\
\colrule
$-0.26$ & $-0.003$ & $1.28f_\pi$  & $1.44$ & 0.036 & $-0.108$   \\
$0.16$ & $-0.003$ & $1.28f_\pi$ & $1.44$ & 0.028 & 0.035   \\
$0.58$ & $-0.003$ & $1.28f_\pi$  & $1.44$ & 0.019 & 0.178   \\
\botrule
\end{tabular} \label{tab:1}
\end{table}

\begin{table}[ht]
\caption{Same as Table \ref{tab:1} except that $\langle
0|\alpha_sG{\tilde G}/(4\pi)|\eta_q\rangle=0.050$ GeV$^3$  and $f_s$ is fixed to trade
for $m_{sg}^2$ as a free parameter.}
\begin{tabular}{c ccc c c c} \toprule
$f_s$ & $r$ & $R$ & $m_{sg}^2$\,(GeV$^2$) & $m_G$\,(GeV) & $\langle
0|{\alpha_s\over 4\pi}G\tilde{G}|\eta_s\rangle$\,(GeV$^3$)
& $\langle 0|{\alpha_s\over 4\pi}G\tilde{G}|g\rangle$\,(GeV$^3$) \\
\colrule
$1.24 f_\pi$ & $0.22$ & $-0.001$ & $-0.009$  & $1.60$ & 0.028 & $0.036$   \\
$1.26 f_\pi$ &
$0.22$ & $-0.003$ & $0.004$  & $1.47$ & 0.028 & 0.036   \\
$1.28 f_\pi$ &
$0.23$ & $-0.005$ & $0.016$  & $1.34$ & 0.028 & 0.038   \\
$1.30 f_\pi$ &
$0.24$ & $-0.007$ & $0.029$  & $1.21$ & 0.028 & 0.040   \\
\colrule
$1.24 f_\pi$ &
$0.12$ & $0.001$ & $-0.054$  & $2.15$ & 0.027 & $0.030$   \\
$1.26 f_\pi$ &
$0.13$ & $-0.001$ & $-0.029$ & $1.84$ & 0.027 & 0.031   \\
$1.28 f_\pi$ &
$0.15$ & $-0.003$ & $-0.005$  & $1.52$ & 0.027 & 0.034   \\
$1.30 f_\pi$ &
$0.24$ & $-0.005$ & $0.018$  & $1.16$ & 0.028 & 0.045   \\
\botrule
\end{tabular} \label{tab:2}
\end{table}
\begin{table}[t]
\caption{Same as Table \ref{tab:2} except $\langle
0|\alpha_sG{\tilde G}/(4\pi)|\eta_q\rangle=0.035$ GeV$^3$.}
\begin{tabular}{c ccc c c c} \toprule
$f_s$ & $r$ & $R$ & $m_{sg}^2$\,(GeV$^2$) & $m_G$\,(GeV) & $\langle
0|{\alpha_s\over 4\pi}G\tilde{G}|\eta_s\rangle$\,(GeV$^3$)
& $\langle 0|{\alpha_s\over 4\pi}G\tilde{G}|g\rangle$\,(GeV$^3$) \\
\colrule
$1.24 f_\pi$ & $0.40$ & $-0.001$ & $-0.009$  & $1.60$ & 0.019 & $0.105$   \\
$1.26 f_\pi$ &
$0.45$ & $-0.003$ & $0.004$  & $1.47$ & 0.020 & 0.113   \\
$1.28 f_\pi$ &
$0.54$ & $-0.005$ & $0.016$  & $1.34$ & 0.021 & 0.126   \\
$1.30 f_\pi$ &
$0.69$ & $-0.007$ & $0.029$  & $1.21$ & 0.022 & 0.148  \\
\colrule
$1.24 f_\pi$ &
$0.27$ & $0.001$ & $-0.054$  & $2.15$ & 0.018 & $0.136$   \\
$1.26 f_\pi$ &
$0.34$ & $-0.001$ & $-0.029$ & $1.84$ & 0.018 & 0.146   \\
$1.28 f_\pi$ &
$0.51$ & $-0.003$ & $-0.005$  & $1.52$ & 0.019 & 0.168   \\
$1.30 f_\pi$ &
$1.18$ & $-0.005$ & $0.018$  & $1.16$ & 0.023 & 0.262   \\
\botrule
\end{tabular} \label{tab:3}
\end{table}

Finally, we explore the impact of $m_{sg}^2$ on our solutions. To
do so, we add $f_s$ as an input so that $m_{sg}^2$ can be
introduced as an unknown. $m_{qg}^2$ is not considered, because
its effect should be very minor as explained before. The results
for the various inputs of $f_s=(1.24$-$1.30) f_\pi$,
$\phi_G=22^\circ$ and $12^\circ$, and $\langle 0|\alpha_sG{\tilde
G}/(4\pi)|\eta_q\rangle=0.050$ (0.035) GeV$^3$ are listed in
Table~\ref{tab:2} (\ref{tab:3}). In the large $N_c$ analysis for
the resolution of the $U(1)$ anomaly
\cite{Witten,Veneziano,Ohta2}, it is the combined contribution
from a contact term and the glueball that cancels the $\eta'$
contribution to give a zero topological susceptibility in full QCD
in the chiral limit. This combined contribution is just the
topological susceptibility $\chi_{\rm quench}$ in the quenched
QCD, which leads to the Witten-Veneziano mass formula $m_{\eta'}^2
= 4 N_F \chi_{\rm quench}/f_{\pi}^2$. $\chi_{\rm quench}$ is
calculated to be $\approx 0.00133$ GeV$^4$ \cite{Debbio}, and the
quenched glueball contributes about \mbox{$- 11\%$} to $\chi_{\rm
quench}$ \cite{YC06}, which makes the contact term to be $\approx
0.00148$ GeV$^4$. It is observed that the anomaly matrix element
$\langle 0|\alpha_sG\tilde{G}/(4\pi)|g\rangle$ is $0.105\, {\rm
GeV}^3$ or larger in Table~\ref{tab:3}. This anomaly matrix
element contributes $\langle
0|\alpha_sG\tilde{G}/(4\pi)|g\rangle^2/(-4m_g^2)\approx -0.00141$
GeV$^4$ to the topological susceptibility for $\langle
0|\alpha_sG\tilde{G}/(4\pi)|g\rangle=0.105$ GeV$^3$ and $m_g =
1.4$ GeV. Namely, the glueball contribution is as large as but
destructive to the contact term. As the glueball contribution and
the contact term already cancel each other to a large extent,
there is no room left for the contact term to cancel the sizable
$\eta$ and $\eta'$ contributions in order to end in a very small
topological susceptibility in full QCD, which has a value $\chi
({\rm full\, QCD}) = -\langle
\bar{\psi}\psi\rangle/(1/m_u+1/m_d+1/m_s) \sim 4 \times 10^{-5}\,
{\rm GeV}^4$ \cite{Leutwyler}. It implies that the anomaly matrix
element $\langle 0|\alpha_sG\tilde{G}/(4\pi)|g\rangle=0.105$
GeV$^3$ is probably too large.

Based on the above reasoning, we believe that $\langle
0|\alpha_sG\tilde{G}/(4\pi)|g\rangle \ge 0.105\, {\rm GeV}^3$ is not
likely to be a viable solution. This criterion would exclude all the
results in Table~\ref{tab:3} with $\langle 0|\alpha_sG{\tilde
G}/(4\pi)|\eta_q\rangle=0.035\, {\rm GeV}^3$ as an input. For
Table~\ref{tab:2}, it is seen that $m_{sg}^2$ and $m_G$ do depend on
$f_s$ sensitively. In some cases, we have $m_G$ as large as 1.84 GeV
and 2.15 GeV, for which $m_{sg}^2$ are negative and large. We cannot
discard these solutions of $m_{sg}^2$ {\it a priori}, but they are
not favored due to their negative values. This issue can be sorted
out, when lattice calculations of $m_{sg}^2$ with dynamical fermions
are available. As $f_s \ge 1.30\, f_{\pi}$, $m_G$ becomes smaller
than 1.2 GeV, where there are no pseudoscalar glueball candidates.
Therefore, if excluding the solutions with large and negative
$m_{sg}^2$,
the range $(1.4 \pm 0.1)$ GeV of the pseudoscalar glueball mass
obtained in Sec.~\ref{glueball} will be more or less respected.

Having studied the higher $O(1/N_c)$ effects and confirmed that they
are small, modulo the uncertainty regarding $m_{sg}^2$, we shall
simply use the typical results in Eq.~(\ref{sol.3})
[Eq.~(\ref{solen})] to obtain the anomaly matrix elements for the
physical states $\eta, \eta'$ and $G$:
\begin{eqnarray}
\langle 0|\alpha_sG{\tilde G}/(4\pi)|\eta\rangle &=&0.026 (0.028)\;\;{\rm GeV}^3\;,\nonumber\\
\langle 0|\alpha_sG{\tilde G}/(4\pi)|\eta'\rangle &=&0.054 (0.057)\;\;{\rm GeV}^3\;,\nonumber\\
\langle 0|\alpha_sG{\tilde G}/(4\pi)|G\rangle &=&-0.059
(-0.041)\;\;{\rm GeV}^3\;.\label{glue}
\end{eqnarray}
It is found that the value of $\langle 0|\alpha_sG{\tilde
G}/(4\pi)|\eta\rangle$ is close to $\sqrt{3/2}f_\eta
m_\eta^2/3\approx 0.021$ GeV$^3$ obtained in \cite{NSVZ}, and
$\langle 0|\alpha_sG{\tilde G}/(4\pi)|\eta'\rangle$ is within a
factor of two from its chiral limit estimated from the topological
susceptibility, i.e., $\langle 0|\alpha_sG{\tilde
G}/(4\pi)|\eta'\rangle =2\sqrt{N_F}\chi/f_{\pi} =0.035\, {\rm
GeV}^3$. Equation~(\ref{glue}) also reveals that $\langle
0|\alpha_sG{\tilde G}/(4\pi)|G\rangle$ is almost the same as that
from the quenched lattice QCD calculation, which gives $|\langle
0|\alpha_sG{\tilde G}/(4\pi)|G\rangle| =(0.06\pm 0.01)\,{\rm GeV}^3$
\cite{YC06}. The fact that $\langle 0|\alpha_sG{\tilde
G}/(4\pi)|\eta'\rangle$ is comparable to $\langle 0|\alpha_sG{\tilde
G}/(4\pi)|G\rangle$, which defies the large $N_c$ scaling in
Eq.~(\ref{largeNc}), implies that the $\eta'$ meson production in
the $J/\psi$ radiative decay may have a branching ratio as large as
that for the pseudoscalar glueball production.

Given the mixing angles, we can predict the widths of the $G\to\gamma\gamma$,
$\ell^+\ell^-$ decays, assuming that they take place through
the quark content \cite{LYF03}. The ratio of the
$G\to\gamma\gamma$ width over the $\pi^0\to\gamma\gamma$ one is
expressed as
\begin{eqnarray}
\frac{\Gamma(G\to\gamma\gamma)}{\Gamma(\pi^0\to\gamma\gamma)}
&=&\frac{1}{9}\left(\frac{m_G}{m_{\pi^0}}\right)^3\left(5\frac{f_\pi}{f_q}
\sin\theta_i\sin\phi_G+\sqrt{2}\frac{f_\pi}{f_s}\cos\theta_i\sin\phi_G\right)^2\;.
\label{R2}
\end{eqnarray}
We have confirmed that both the parameter sets in \cite{KLOE} and
\cite{EN07} give the $\eta, \eta'\to\gamma\gamma$ widths in
agreement with the data $\Gamma(\eta\to\gamma\gamma)\approx 0.51$
keV and $\Gamma(\eta'\to\gamma\gamma)\approx 4.28$ keV \cite{PDG},
by considering the similar ratios for the $\eta, \eta'$ mesons. The
parameter set in \cite{KLOE} (\cite{EN07}) leads to a ratio 387
(83.3) in Eq.~(\ref{R2}), i.e., the decay width
$\Gamma(G\to\gamma\gamma)=3 (0.6)$ keV for
$\Gamma(\pi^0\to\gamma\gamma)=7.7$ eV \cite{PDG}. If $\eta(1405)$ is
a pseudoscalar glueball, we predict the branching ratio ${\cal
B}(\eta(1405)\to\gamma\gamma)=6 (1)\times 10^{-5}$, i.e., an order
of $10^{-5}$  for the total decay width $\Gamma(\eta(1405))=51$ MeV
\cite{PDG}. The above result can be confronted with future
experimental data. The ``stickiness" $S$ has been proposed to be a
useful quantity for identifying a glueball rich state
\cite{Chanowitz}, which is defined as the ratio of $\Gamma(J/\psi\to
\gamma G)$ to $\Gamma(G\to \gamma\gamma)$ with the phase space
factors taken out. Combining our predictions for the pseudoscalar
glueball production and decay, we obtain $S=18$-80 for $G$, which is
much larger than $S=1$ as defined for the $\eta$ meson.

For the $G\to\ell^+\ell^-$ decays, we calibrate their widths using
the available $\pi^0\to e^+e^-$ and $\eta\to\mu^+\mu^-$ data:
\begin{eqnarray}
\frac{\Gamma(G\to e^+e^-)}{\Gamma(\pi^0\to e^+e^-)}
&=&\frac{1}{9}\left(\frac{m_G}{m_{\pi^0}}\right)^3\left(5\frac{f_\pi}{f_q}
\sin\theta_i\sin\phi_G+\sqrt{2}\frac{f_\pi}{f_s}\cos\theta_i\sin\phi_G\right)^2\;,
\nonumber\\
\frac{\Gamma(G\to\mu^+\mu^-)}{\Gamma(\eta\to\mu^+\mu^-)}
&=&\left(\frac{m_G}{m_{\eta}}\right)^3\left(5\frac{f_\pi}{f_q}
\sin\theta_i\sin\phi_G+\sqrt{2}\frac{f_\pi}{f_s}\cos\theta_i\sin\phi_G\right)^2
\nonumber\\
& &\times \left[5\frac{f_\pi}{f_q}
\left(\cos\phi+\sin\theta\sin\theta_i\Delta_G\right)
-\sqrt{2}\frac{f_\pi}{f_s}\left(\sin\phi+\sin\theta\cos\theta_i\Delta_G
\right)\right]^{-2}\;.
\end{eqnarray}
For $\Gamma(\pi^0\to e^+e^-)=4.8\times 10^{-7}$ eV \cite{PDG}, we
obtain $\Gamma(G\to e^+e^-)=1.9 (0.4)\times 10^{-4}$ eV using the
parameter set from \cite{KLOE} (\cite{EN07}). For
$\Gamma(\eta\to\mu^+\mu^-)=7.5\times 10^{-3}$ eV \cite{PDG}, we have
$\Gamma(G\to\mu^+\mu^-)=4.0(1.0)\times 10^{-2}$ eV. If
$\eta(1405)$ is a pseudoscalar glueball, the above predictions
correspond to the branching ratios ${\cal B}(\eta(1405)\to
e^+e^-)=4 (0.8)\times 10^{-12}$ and ${\cal B}(\eta(1405)\to
\mu^+\mu^-)=8 (2)\times 10^{-10}$, which would be quite a
challenge to observe experimentally.

\section{CONCLUSION}

In this paper, we have formulated the $\eta$-$\eta'$-$G$ mixing scheme
via the vacuum to meson transition matrix elements for the
anomalous Ward identity. The extension to include the glueball
mixing with the flavor-singlet $\eta_1$ is a generalization of the
FKS scheme for the $\eta$-$\eta'$ mixing \cite{FKS}. Therefore,
only one extra angle $\phi_G$ for the mixing of the glueball state
$g$ and $\eta_1$ is introduced in addition to the angle $\phi$ in
the FKS scheme. We have explained the different parameter
extractions from the same set of $\eta, \eta'$ meson data in
\cite{KLOE} and \cite{EN07}, which give an idea of the
uncertainties contained in the inputs. The obtained pseudoscalar
glueball mass $m_G$ around 1.4 GeV is much lower than the results
from quenched lattice QCD ($>2.0$ GeV). It has been examined that
our solution for $m_G$ depends weakly on the ratio of the decay
constants $f_s/f_q$ in the favored phenomenological range and is stable
against the variation of $\phi_G$ and the higher $O(1/N_c)$ corrections.

There may not exist a unique feature which tells a glueball apart
from a quark-antiquark state. We need to combine information from
$J/\psi$ radiative decays, hadronic decays, as well as
$\gamma\gamma$ and leptonic decays as advocated in \cite{LLI89}. The
comparison of our solutions with the available data suggests that
$\eta(1405)$, which is copiously produced in the $J/\psi$ radiative
decay but has not been seen in the $\gamma\gamma$
reaction, is a strong pseudoscalar glueball candidate.
The anomaly matrix elements $\langle 0|\alpha_sG{\tilde
G}/(4\pi)|\eta'\rangle$ and $\langle 0|\alpha_sG{\tilde
G}/(4\pi)|G\rangle$ in Eq.~(\ref{glue}) are in reasonable agreement
with those estimated from the topological susceptibility and quenched
lattice calculation.  According to our
analysis, the $\eta(1405)\to \gamma\gamma$ decay width is
0.6-3 keV, and the leptonic decays $\eta(1405)\to\ell^+\ell^-$ are
very small. Both predictions can be confronted with future experiments.

\vskip 1.0cm This work was supported by the National Science Council
of R.O.C. under the Grant Nos. NSC-95-2112-M-050-MY3,
NSC96-2112-M-001-003 and by the National Center for Theoretical
Sciences of R.O.C. It is also partially supported by U.S. DOE
grant no. DE-FG05-84ER40154. HYC and HNL acknowledge the hospitality of R.
Fleischer during their visit to the CERN Theory Institute in June
2008, and of Y.L. Wu during their visit to the Kavli Institute for
Theoretical Physics, China, where part of this work was done. KFL would
like to acknowledge the hospitality of Academia Sinica during his
visit when this work was initiated.

\end{document}